\documentclass[showpacs,aps,pra,twocolumn]{revtex4}
\usepackage{amsmath,graphicx,bbm,mathrsfs,amssymb,pst-all,bm,color}

\begin{document}

\title{A quantum algorithm for solving some discrete mathematical problems
by probing their energy spectra}
\author{Hefeng Wang$^1$\footnote{Correspondence to wanghf@mail.xjtu.edu.cn}}
\author{Heng Fan$^2$}
\author{Fuli Li$^1$}
\affiliation{$^{1}$Department of Applied Physics, Xi'an Jiaotong University, Xi'an
710049, China\\
$^{2}$Institute of Physics, Chinese Academy of Science, Beijing 100190, China}

\begin{abstract}
When a probe qubit is coupled to a quantum register that represents a
physical system, the probe qubit will exhibit a dynamical response only when
it is resonant with a transition in the system. Using this principle, we
propose a quantum algorithm for solving discrete mathematical problems based
on the circuit model. Our algorithm has favorable scaling properties in
solving some discrete mathematical problems. \noindent
\end{abstract}

\maketitle

\section{Introduction}

A lot of progress has been made in the field of quantum computation since
the discovery of Shor's factoring algorithm~\cite{shor} and Grover's search
algorithm~\cite{grover}. Quantum computing offers an increase in calculation
speed for a number of problems~\cite{childs,nori}. It has been suggested
that for a fairly general class of quantum systems, especially discrete
systems, an exponential increase in speed can be achieved using quantum
simulators~\cite{sl}.

In Ref.~\cite{farhi1,farhi2}, Farhi and coworkers developed a quantum
adiabatic algorithm~(QAA) for solving a discrete mathematical problem, the $%
3 $-bit exact cover problem~(EC$3$). In this algorithm, one starts from an
initial Hamiltonian, $H_{B}$, and using its ground state as the initial
state, and $H_{B}$ evolves to a final Hamiltonian $H_{P}$, whose ground
state is the solution to the EC$3$ problem. The system evolves from the
initial state to the ground state of $H_{P}$.

In this paper, we propose an alternative quantum algorithm for solving some
discrete mathematical problems based on the circuit model. Our approach can
be applied to any satisfiability problem in principle, here we demonstrate
it with a particular instance of the exact cover problem, EC$3$.

The EC$3$ problem on a quantum computer can be formulated as follows~\cite%
{farhi1,farhi2}: the $3$-bit instance of satisfiability is a Boolean formula
with $M$ clauses
\begin{equation}
C_{1}\wedge C_{2}\wedge \cdots \wedge C_{M}\text{,}
\end{equation}%
where each clause $C_{l}$ is true or false depending on the values of a
subset of the $n$ bits, and each clause contains three bits. The clause is
true if and only if one of the three bits is $1$ and the other two are $0$.
The task is to determine whether one~(or more) of the $2^{n}$ assignments
satisfies all of the clauses, that is, makes formula~($1$) true, and find
the assignment(s) if it exists. Let $i_{C}$, $j_{C}$ and $k_{C}$ be the $3$
bits associated with clause \textit{C}. For each clause \textit{C}, we
define an \textquotedblleft energy\textquotedblright\ function
\begin{equation}
h_{C}(z_{i_{C}},z_{j_{C}},z_{k_{C}})\!=\!\Bigg\lbrace%
\begin{array}{c}
\!\!\!\!0,\,\,\mathrm{if}\,(z_{i_{C}},z_{j_{C}},z_{k_{C}})\ \mathrm{%
satisfies\ clause\ \emph{C}} \\
\!\!\!\hskip.0003in1,\,\,\mathrm{if}\,(z_{i_{C}},z_{j_{C}},z_{k_{C}})\
\mathrm{violates\ clause\ \emph{C}}\,%
\end{array}%
\end{equation}%
Then
\begin{equation}
H_{P,C}|z_{1}z_{2}\cdots z_{n}\rangle =h_{C}\left(
z_{i_{C}},z_{j_{C}},z_{k_{C}}\right) |z_{1}z_{2}\cdots z_{n}\rangle ,
\end{equation}%
where $|z_{j}\rangle $ is the $j$-th bit and has a value $0$ or $1$. Define
\begin{equation}
H_{P}=\sum_{C}H_{P,C}.
\end{equation}%
$H_{P}|\psi \rangle =0$, if and only if $|\psi \rangle $ is a state of the
form $|z_{1}z_{2}\cdots z_{n}\rangle $, where the bit string $%
z_{1}z_{2}\cdots z_{n}$ satisfies all of the clauses, or a superposition of
such states. If formula~($1$) has no satisfying assignments, the ground
state~(or states) of $H_{P}$ corresponds to the assignment~(or assignments)
that violates the fewest clauses. The computational basis of $%
|z_{1}z_{2}\cdots z_{n}\rangle $ is of dimension $N=2^{n}$.

\section{The algorithm}

Our algorithm for solving the EC$3$ problem is described below.

First, we construct a Hamiltonian $\widetilde{H}$ with the form:
\begin{equation}
\widetilde{H}=\left(
\begin{array}{cc}
-I_{N} & 0 \\
0 & H_{P}%
\end{array}%
\right) ,
\end{equation}%
where $I_{N}$ is the $N$-dimensional identity operator, and $\widetilde{H}$
acts on the state space of a ($n+1$)-qubit quantum register $R_{S}$, which
contains one ancilla qubit and $n$ qubits that represents the system. Then
we let a probe qubit couple to $R_{S}$, and design a Hamiltonian $H$ for the
whole system of the form
\begin{equation}
H=\frac{1}{2}\omega \sigma _{z}\otimes I_{2}^{\otimes \left( n+1\right)
}+I_{2}\otimes \widetilde{H}+c\sigma _{x}\otimes A,
\end{equation}%
where $I_{2}$ is the two-dimensional identity operator. In Eq.~($6$), the
first term is the Hamiltonian of the probe qubit, the second term is the
Hamiltonian of the quantum register $R_{S}$, and the third term describes
the interaction between the probe qubit and $R_{S}$. Here, $\omega $ is the
frequency of the probe qubit~($\hbar =1$), and $c$ is the coupling strength
between the probe qubit and $R_{S}$, whereas $\sigma _{x}$ and $\sigma _{z}$
are Pauli matrices. The operator $A$ acts on the state space of $\widetilde{H%
}$ and plays the role of an excitation operator. This operator contains $N$
terms, which provide all possible terms that excite the system from the
subspace of $-I_{N}$ to the subspace of $H_{P}$,
\begin{equation}
A=\frac{1}{\sqrt{N}}(A_{1}+A_{2}+\cdots +A_{N}),
\end{equation}%
where
\begin{eqnarray}
A_{1} &=&\sigma _{x}\otimes I_{2}\otimes I_{2}\otimes \cdots \otimes I_{2},
\notag \\
A_{2} &=&\sigma _{x}\otimes I_{2}\otimes \cdots \otimes I_{2}\otimes \sigma
_{x},  \notag \\
&&\cdots ,  \notag \\
A_{n+1} &=&\sigma _{x}\otimes \sigma _{x}\otimes I_{2}\otimes I_{2}\otimes
\cdots \otimes I_{2},  \notag \\
A_{n+2} &=&\sigma _{x}\otimes I_{2}\otimes \cdots \otimes I_{2}\otimes
\sigma _{x}\otimes \sigma _{x},  \notag \\
A_{n+3} &=&\sigma _{x}\otimes I_{2}\otimes \cdots \otimes \sigma _{x}\otimes
I_{2}\otimes \sigma _{x},  \notag \\
&&\cdots ,  \notag \\
A_{N} &=&\sigma _{x}\otimes \sigma _{x}\otimes \cdots \otimes \sigma
_{x}\otimes \sigma _{x}.
\end{eqnarray}%
Although the operator $A$ has an exponentially large number of terms, it can
be implemented at a polynomial cost as
\begin{equation}
A=\sigma _{x}\otimes \left[ \frac{1}{\sqrt{2}}(I_{2}+\sigma _{x})\right]
^{\otimes n}.
\end{equation}

We prepare $R_{S}$ in state
\begin{equation}
|\Psi _{0}\rangle =\frac{1}{\sqrt{N}}\sum_{j=1}^{N}|\varphi _{j}\rangle =%
\frac{1}{\sqrt{N}}\sum_{j=1}^{N}|0\rangle \otimes |j-1\rangle,
\end{equation}%
as the reference state, where $|\varphi _{j}\rangle =$ $|0\rangle \otimes
|j-1\rangle $ and $|j-1\rangle $ are the states of the computational basis
in the subspace of $I_{N}$, and $|0\rangle =\left(
\begin{array}{c}
1 \\
0%
\end{array}%
\right) $. This is achieved by initializing $R_{S}$ in state $|0\rangle
^{\otimes \left( n+1\right) }$ and applying the operator $I_{2}\otimes
H_{d}^{\otimes n}$, where $H_{d}$ is the Hadamard gate. The states $%
|\varphi_{j}\rangle $ are all eigenstates of $\widetilde{H}$ with
eigenvalues of $-1$. Therefore, the reference state $|\Psi _{0}\rangle $ is
also an eigenstate of $\widetilde{H}$\ with eigenvalue $E_{0}=-1$. We set
the frequency of the probe qubit $\omega =1$ and run the circuit in Fig.~$1$%
. We repeat this procedure many times until the probe qubit decays to its
ground state.
\begin{figure}[tbp]
\includegraphics[width=0.9\columnwidth, clip]{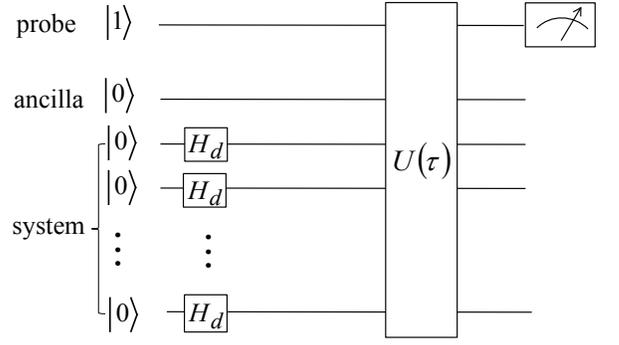}
\caption{Quantum circuit for solving the exact cover problem EC$3$. $(n+2)$
qubits are required to solve an $n$-bit EC$3$ problem. The first line
represents a probe qubit, and the next $n+1$ lines represent the quantum
register $R_{S}$. $H_{d}$ represents the Hadamard gate, and $U(\protect\tau %
) $ is the time evolution operator for the Hamiltonian given in Eq.~($6$).}
\end{figure}

The algorithm procedure is summarized as follows: $\left( i\right) $ prepare
the first qubit in state $|1\rangle $ and the next ($n+1$)-qubit quantum
register $R_{S}$ in state $|0\rangle ^{\otimes \left( n+1\right) }$; $\left(
ii\right) $ apply operator $I_{2}\otimes H_{d}^{\otimes n}$ on $R_{S}$,
therefore $R_{S}$ is transformed into state $|\Psi _{0}\rangle $ as shown in
Eq.~($10$); $\left( iii\right) $ implement the time evolution operator $%
U(\tau )=\exp \left( -iH\tau \right) $ where $H$ is given in Eq.~($6$); $%
\left( iv\right) $ read out the state of the probe qubit; $\left( v\right) $
repeat steps $\left( i\right) $ -- $\left( iv\right) $ until a decay of the
probe qubit to its ground state is observed. The quantum circuit for steps $%
\left( i\right) $ -- $\left( iv\right) $ is shown in Fig.~$1$.

\section{Efficiency of the algorithm}

In our algorithm, $(n+2)$ qubits are required for solving an $n$-bit EC$3$
problem on a quantum computer, which increases linearly with the size of the
problem. In the following paragraphs, we discuss the efficiency of the
algorithm.

The efficiency of the algorithm is defined as the number of times that the
circuit in Fig.~$1$ must be run to observe a decay of the probe qubit. The
number of times that the circuit must be run must be at least proportional
to $1/P_{\text{decay}}$. The decay probability of a probe qubit that coupled
to a system has been discussed in Ref.~\cite{wan}. In our case, we prepare $%
R_{S}$ in state $|\Psi _{0}\rangle $ and only consider the excitation from
the subspace of $-I_{N}$ to the subspace of $H_{P}$. Therefore the decay
probability of the probe qubit becomes:
\begin{equation}
P_{\text{decay}}=\sin ^{2}\left( \frac{\Omega _{0j}\tau }{2}\right) \frac{%
Q_{0j}^{2}}{Q_{0j}^{2}+\left( E_{j}-E_{0}-\omega \right) ^{2}},
\end{equation}%
where
\begin{equation}
Q_{0j}=2c|\langle \Psi _{j}|A|\Psi _{0}\rangle |,
\end{equation}%
and
\begin{equation}
\Omega _{0j}=\sqrt{Q_{0j}^{2}+\left( E_{j}-E_{0}-\omega \right) ^{2}},
\end{equation}%
$|\Psi _{j}\rangle $~($j=1,2,\ldots ,N$) is the $j$-th energy eigenstate and
$E_{j}$ is the corresponding eigenenergy of $H_{P}$, and according to Eq.~($%
4 $), $E_{j}$ are discrete integers. Eq.~($11$) describes Rabi-oscillation
dynamics, in which the quantum register $R_{S}$ and the probe qubit exchange
an excitation; $R_{S}$ is excited from state $|\Psi _{0}\rangle $ to state $%
|\Psi _{j}\rangle $.

If a solution to the EC$3$ problem exists, the excitation frequency between
the reference state $|\Psi_{0}\rangle $ and the state $|\Psi _{1}\rangle $
with eigenvalue $E_{1}=0$, which contains all solutions to the problem, is $%
1 $. With the probe qubit frequency being set to $\omega =1$ and assuming
there exists a solution to the problem, the decay probability of the probe
qubit becomes
\begin{equation}
P_{\text{decay}}=\sin ^{2}\left( \frac{Q_{01}\tau }{2}\right),
\end{equation}%
where
\begin{eqnarray}
Q_{01} &=&2c|\langle \Psi _{1}|A|\Psi _{0}\rangle |  \notag \\
&=&2c\sum_{j=1}^{N}\sum_{i=1}^{m}\frac{1}{\sqrt{Nm}}|\langle 1|\langle \mu
_{i}|A|\varphi _{j}\rangle |  \notag \\
&=&2cNm\frac{1}{\sqrt{Nm}}\frac{1}{\sqrt{N}}  \notag \\
&=&2c\sqrt{m}.
\end{eqnarray}%
This term describes the summation over all excitation channels in the
Rabi-oscillation. Here, $|\Psi _{1}\rangle =\sum_{i=1}^{m}|1\rangle |\mu
_{i}\rangle /\sqrt{m}$ encodes all solutions to the EC$3$ problem, which is
a superposition of all $m$ assignments that satisfy all of the clauses. In
addition, $|\mu _{i}\rangle $ are these assignments, which are the basis
states of $H_{P}$ with eigenvalues $0$ and degeneracy $m$. Here, although $1/%
\sqrt{N}$ appears in both the operator $A$ and the reference state $|\Psi
_{0}\rangle $, there are $N$ terms in the excitation operator $A$, which
provide $N$ excitation channels. For each state $|\mu _{i}\rangle $, there
are $N$ excitation channels connecting it to the reference state $|\Psi
_{0}\rangle $. Considering the degeneracy of the states $|\mu _{i}\rangle $%
,these channels contribute a factor of $Nm$ to the whole term. From Eq.~($15$%
), we can see that as $m$ increases, there are more excitation channels, and
the period of $P_{\text{decay}}$ decreases. By knowing $Q_{01}$, we can
choose a specific evolution time $\tau $ considering the degeneracy of the
ground state of the problem, such that the probe qubit has a high decay
probability and therefore the algorithm has a high efficiency. By attempting
to guess $m$, one can quickly identify the optimal evolution time $\tau $
for implementing the algorithm.

From Eq.~($14$) and Eq.~($15$) above, we can see that the decay probability,
and therefore the efficiency of the algorithm, depends on the coupling
strength $c$ and the evolution time $\tau $. In general, we need to set $c$
to be small so that we have weak system-probe coupling. The evolution time $%
\tau $ should be large, such that the change of the system is clear and one
obtains a high decay probability.

When we set $\omega =1$, the coupling between the reference state and all
other states, except the state with an eigenvalue equal to zero, also
contributes to the decay probability of the probe qubit, and therefore
introduces an error in $P_{\text{decay}}$. We now evaluate this error, $P_{%
\text{decay}}^{\text{err}}$.
\begin{eqnarray}
P_{\text{decay}}^{\text{err}} &=&\sum_{j=2}^{N}\sin ^{2}\left( \frac{\Omega
_{0j}\tau }{2}\right) \frac{Q_{0j}^{2}}{Q_{0j}^{2}+\left( E_{j}-E_{0}-\omega
\right) ^{2}}  \notag \\
&<&\sum_{j=2}^{N}\frac{Q_{0j}^{2}}{E_{j}{}^{2}}  \notag \\
&\leq &\sum_{j^{\prime }=2}^{N_{m}}\frac{4c^{2}m_{j}}{\left( j^{\prime
}-1\right) ^{2}}  \notag \\
&<&4c^{2}m_{0}\frac{\pi ^{2}}{6}=\frac{2}{3}m_{0}\pi ^{2}c^{2},
\end{eqnarray}%
where $E_{0}=-1$ is the eigenvalue of the reference state and $m_{j}$
represents the degeneracy of the assignments with eigenvalue $E_{j}$. $N_{m}$
is the eigenenergy of the highest energy level of the problem, and as $%
N_{m}\rightarrow \infty $, $\sum_{j^{\prime }=2}^{N_{m}}\frac{1}{\left(
j^{\prime }-1\right) ^{2}}=\frac{\pi ^{2}}{6}$. $m_{0}$ is the maximum of $%
m_{j}$. When there is no energy level with exponentially large degeneracy,
the term $2m_{0}\pi ^{2}c^{2}/3$ can be small because $c$ can be set small,
but not exponentially small. We can then constrain this error to be very
small. In this case, the algorithm can be completed in a finite time $\tau $.

\section{Implementation of the algorithm}

We now discuss the implementation of the algorithm. In the algorithm, we
must implement the time evolution operator $U(\tau )=\exp \left( -iH\tau
\right)$. In the Hamiltonian $H$, as shown in Eq.~($6$), the first two terms
commute with each other, while they do not commute with the third term. The
operator $U(\tau )$ can be implemented using the procedure of quantum
simulation based on the Trotter-Suzuki formula~\cite{nc}:%
\begin{equation}
U(\tau )\!=\!\left[ e^{-i\left( \frac{1}{2}\omega \sigma _{z}+\widetilde{H}%
\right) \tau /L}e^{-i\left( c\sigma _{x}\otimes A\right) \tau /L}\right]
^{L}\!+\!O\!\left( \frac{1}{L}\right) ,
\end{equation}%
where $L$ does not depend on the size of the problem. $L$ can be made
sufficiently large such that the error is bounded by some threshold~\cite{sl}%
.
\begin{figure}[tbp]
\includegraphics[width=0.9\columnwidth, clip]{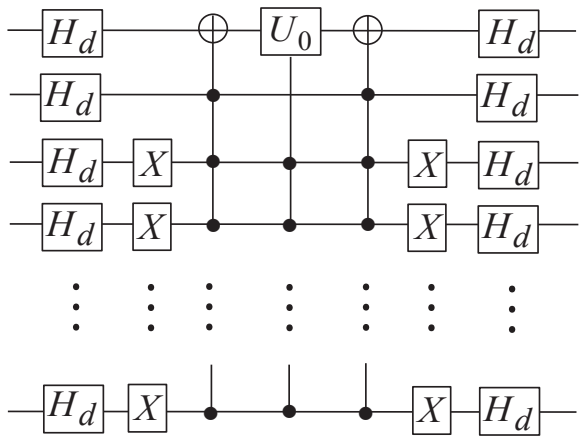}
\caption{Quantum circuit for implementing the unitary operator $e^{-i(c%
\protect\sigma_{x}\otimes A)\protect\tau /L}$, where operator $A$ is given
in Eq.~($9$). $H_{d}$ is the Hadamard gate and $U_{0}=e^{i\protect\sqrt{2^{n}%
}c\protect\tau /L\protect\sigma _{z}}$, where $\protect\sigma_{z}$ are the
Pauli matrices.}
\end{figure}

In Eq.~($17$), the unitary operator $e^{-i\left( \frac{1}{2}\omega \sigma
_{z}+\widetilde{H}\right) \tau /L}$ is diagonal and can be efficiently
implemented. For the unitary operator $e^{-i\left( c\sigma _{x}\otimes
A\right) \tau /L}$, the Hamiltonian $c\sigma _{x}\otimes A$ involves a
many-body interaction. In Ref.~\cite{bravyi}, it was shown that a many-body
interaction Hamiltonian can be efficiently simulated by a Hamiltonian with
two-body interactions. The unitary operator $e^{-i(c\sigma _{x}\otimes
A)\tau /L}$ can be implemented using the circuit shown in Fig.~$2$. In the
circuit, the $\left( n+1\right) $ and $n$-qubit controlled unitary operators
can be efficiently implemented with $O \left( n^{2}\right) $ elementary
gates~\cite{barenco}.

The second term in the Hamiltonian $H$, $\widetilde{H}$, as shown in Eq.~($5$%
), can be seen as a controlled-$H_{P}$~(C-$H_{P}$) operation. The operation
that calculates $H_{P}$ can be taken as an oracle. In our algorithm, this
oracle is entangled with the probe qubit. The number of times that the C-$%
H_{P}$ oracle is implemented is $L$. Therefore, the implementation of $%
U(\tau )$ scales polynomially with the size of the problem.

\section{Example: solving an $8$-bit EC$3$ problem}

In the following, we present an example to demonstrate an application of the
algorithm.

Considering an $8$-bit EC$3$ problem, we chose three cases of the $3$-bit
sets of clauses, which have one, two, and four satisfying assignments for
the problem, respectively. As discussed above, $10$ qubits are required to
solve this problem using our algorithm. We set the coupling coefficient $%
c=0.002$ and run the algorithm for different evolution times $\tau $. As the
probe qubit decays to the state $|0\rangle $, the state of the last $8$
qubits of the quantum register $R_{S}$ encodes all of the solutions to the
above EC$3$ problem. Case $i)$: the $3$-bit sets are \{$1,2,8$\}, \{$2,3,6$%
\}, \{$2,3,7$\}, \{$2,4,5$\}, \{$2,5,6$\}, and \{$3,5,8$\}. The solution to
the $8$-bit EC$3$ problem for this case is $|00010111\rangle $. Case $ii)$:
the $3$-bit sets are chosen to be \{$1,4,5$\}, \{$1,7,8$\}, \{$2,4,8$\}, \{$%
2,7,8$\}, \{$4,5,8$\}, and \{$5,6,7$\}. The solution to the problem for this
case is $|00010010\rangle $ and $|00110010\rangle $. Case $iii)$: the $3$%
-bit sets are \{$1,3,5$\}, \{$1,6,8$\}, \{$2,4,6$\}, \{$2,6,8$\}, and \{$%
4,5,7$\}. The solution to the problem for this case is $|00001100\rangle $, $%
|00100110\rangle $, $|00110001\rangle $, and $|11000010\rangle $.
\begin{figure}[tbp]
\includegraphics[width=0.9\columnwidth, clip]{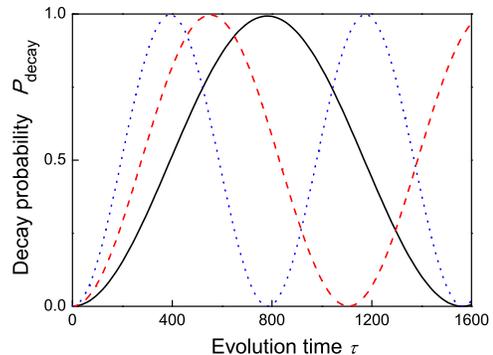}
\caption{(Color online) Decay probability of the probe qubit vs. evolution
time $\protect\tau$. The frequency of the probe qubit $\protect\omega =1$
and the coupling strength $c=0.002$. The black solid line shows the result
for case $i$): the $3$-bit sets are \{$1,2,8$\}, \{$2,3,6$\}, \{$2,3,7$\}, \{%
$2,4,5$\}, \{$2,5,6$\}, and \{$3,5,8$\}; the red dashed line shows the
result for case $ii$): the $3$-bit sets are \{$1,4,5$\}, \{$1,7,8$\}, \{$%
2,4,8$\}, \{$2,7,8$\}, \{$4,5,8$\}, and \{$5,6,7$\}; and the blue dotted
line shows the result for case $iii$): the $3$-bit sets are \{$1,3,5$\}, \{$%
1,6,8$\}, \{$2,4,6$\}, \{$2,6,8$\}, and \{$4,5,7$\}. The simulated results
fit exactly with the analytical results predicted in Eq.~($14$). Atomic
units are used in the figure.}
\end{figure}

In Fig.~$3$, we show the variation of simulated $P_{\text{decay}}$ with the
evolution time $\tau $ for the above three cases. The black solid line, the
red dashed line, and the blue dotted line show the results for cases $i)$, $%
ii)$ and $iii)$, respectively. In the above three cases, the decay
probability almost reaches unity at $\tau =800$, $\tau =550$, and $\tau =400$%
, respectively. This result shows that the algorithm can be run efficiently.
From the figure we can also see that as the number of satisfying assignments
increases, the period of $P_{\text{decay}}$ decreases. The simulated $P_{%
\text{decay}}$ fits exactly with the analytical results predicted by Eq.~($%
14 $). In the case in which there is no solution to the problem, if we still
set the probe qubit frequency $\omega =1$, then $P_{\text{decay}}$
approaches zero.

\section{Discussion}

We have developed a quantum algorithm for solving a specific discrete
mathematical problem, the EC$3$ problem. In our algorithm for solving the EC$%
3$ problem, we construct a Hamiltonian that contains the Hamiltonian of the
problem and a Hamiltonian with the same dimension $N$ as the problem, whose
eigenstates have degeneracy $N$ and eigenvalues~(in our case, $-1$) lower
than the smallest eigenvalue of the problem. The second Hamiltonian is used
as a reference point. If a solution to the problem exists, we will observe
the decay of the probe qubit at the excitation frequency~(in our case, $1$)
between the reference state and the eigenstates with an eigenvalue of zero.
In this case, the system register evolves to its ground state, which is a
superposition state that encodes the solution to the EC$3$ problem. If there
is no solution to the problem, we can increase the frequency of the probe
qubit discretely~(because the energy function of the problem is discrete).
The first frequency at which the probe qubit decays indicates the ground
state of the system which encodes the assignment that violates the smallest
number of clauses.

In our algorithm, one can determine whether there exists a solution to the
problem immediately by performing a measurement on the probe qubit. If the
probe qubit decays to its ground state, it indicates that a solution to the
problem exists, otherwise the problem has no solution. The last $n$ qubits
of the register $R_{S}$ encodes the solution to the problem if one observes
a decay of the probe qubit. The solution can be in a superposition state of
the multiple bit strings that satisfy the Boolean formula. And these bit
strings can be obtained by employing the quantum state tomography.

Our algorithm can be applied to a large number of discrete mathematical
problems, such as some combinatorial optimization problems. The procedure is
similar: map the discrete mathematical problem on a quantum computer,
therefore the problem is transferred to looking for the eigenstates with the
lowest eigenenergy of the quantum system; Construct a $(n+1)$-qubit quantum
register, which contains one ancilla qubit and $n$ qubits that represents
the system. Then couple this quantum register with a probe qubit, which is
used for probing the energy spectra of the system. And the whole system is
driven by the Hamiltonian as shown in Eq.~($6$). By varying the frequency of
the probe qubit discretely, the ground state or any desired eigenstates of
the system can be found, which in general, encodes the information of the
solution to the problem.

\begin{acknowledgements}
We are grateful to Sahel Ashhab and Franco Nori for insightful discussions
and critical reading of the manuscript. We thank L.-A. Wu for helpful
discussions. This work was supported by the National Nature Science
Foundation of China~(Grants No.~11275145, No.~11305120 and No.~11074199), \textquotedblleft the Fundamental Research Funds for the Central Universities\textquotedblright\ of China, the
National Basic Research Program of China~(Grant No.~2010CB923102 and
2010CB922904), and the Special Prophase Project on the National Basic Research
Program of China~(Grant No.~2011CB311807).
\end{acknowledgements}

\end{document}